\documentclass[11pt,a4paper,english,superscriptaddress,aps,preprint]{revtex4}
\usepackage{amsmath}
\usepackage{amssymb}
\makeatletter
\usepackage{babel}

\newcommand{\vk}{\vec{k}}
\newcommand{\e}{\epsilon}

\usepackage{hyperref}

\begin{document}

\title{Lorentz symmetry breaking in the noncommutative Wess-Zumino 
model: One loop corrections}

\author{A. F. Ferrari}
\affiliation{Instituto de F\'{\i}sica, Universidade de S\~{a}o Paulo, 
Caixa
Postal 66318, 05315-970, S\~{a}o Paulo - SP, Brazil}
\email{alysson,mgomes@fma.if.usp.br}

\author{H. O. Girotti}
\affiliation{Instituto de F\'{\i}sica, Universidade Federal do Rio 
Grande do
Sul, Caixa Postal 15051, 91501-970 - Porto Alegre, RS, Brazil}
\email{hgirotti@if.ufrgs.br}

\author{M. Gomes}
\affiliation{Instituto de F\'{\i}sica, Universidade de S\~{a}o Paulo, 
Caixa
Postal 66318, 05315-970, S\~{a}o Paulo - SP, Brazil}
\email{mgomes@fma.if.usp.br}

\begin{abstract}
In this paper we deal with the issue of Lorentz symmetry breaking in
quantum field theories formulated in a noncommutative space-time. We
show that, unlike in some recent analysis of quantum gravity effects,
supersymmetry does not protect the theory from the large Lorentz
violating effects arising from the loop corrections. We take advantage
of the noncommutative Wess-Zumino model to illustrate this point.
\end{abstract}

\maketitle

It has been common belief that the possible breaking of Lorentz 
symmetry induced by quantum gravity at the Planck scale is suppressed by 
enormous ratios at low energies. However, recently, Collins et 
al.~\cite{Collins} have argued that this is not generally the case. Indeed, using 
the the Yukawa theory as an example, it was show that quadratically 
divergent one-loop corrections to the pion self-energy translate a breaking 
of Lorentz symmetry at the Planck scale into an observable violation of 
Lorentz invariance at low energies. Needless to say, this result is 
inconsistent with experimental constraints. One way to circumvent this 
problem is to reestablish the Lorentz symmetry at low energies by adding 
appropriately adjusted Lagrangian counterterms or, what amounts to the 
same thing, a Lorentz fine tuning mechanism appears to be required to 
conciliate theory with experiment. 

This Lorentz fine tuning problem is very general and it is present in commutative as well as in noncommutative field theories. With respect to the last ones, the mixing of ultraviolet (UV) and infrared (IR) singularities promoted by the noncommutativity of the spacetime coordinates is known to generate, in non-supersymmetric models, Lorentz violating operators which are too large to be consistent with low-energy tests of Lorentz symmetry~\cite{anisimov}. In view of these considerations, it was even suggested that (Lorentz violating) noncommutative field theories are ruled out by the currently available tests of Lorentz invariance~\cite{Collins}.

Nevertheless, one might expect to avoid the need of fine-tuned counterterms to protect Lorentz symmetry at low energies if the theory under consideration is supersymmetric. In the context of commutative field theories, this was shown to be the case in Ref.~\cite{Jain}, using the Wess-Zumino model as a prototype. As far as noncommutative models are concerned, supersymmetry is well known to avoid the problem of nonintegrable UV/IR infrared singularities~\cite{Mat,Girotti3}, thus allowing the construction of a consistent perturbative expansion of these models. 

It it natural, therefore, to investigate whether supersymmetry can also prevent the Lorentz fine tuning problem in noncommutative field theories. This paper is dedicated to clarify this question.  
We use, as arena, the noncommutative Wess-Zumino (WZ) model since, for the time being, it is the only known $(3 + 1)$-dimensional noncommutative field theory which has been 
proved to be renormalizable to all orders of perturbation 
theory~\cite{Girotti3,Bichl}. Furthermore, the fact that all ultraviolet 
divergences are logarithmic secures that the UV/IR mixing does not give rise 
to nonintegrable infrared divergences. We focus on the 
renormalized two-point vertex function of one of the component fields. The 
one-loop correction to this vertex function is contributed by planar as well 
as by nonplanar integrals. The planar integrals are, of course, 
Lorentz invariant, while the nonplanar ones give, in spite of being 
ultraviolet finite, a non-vanishing contribution to the quantity ($\xi$) 
introduced by Collins et al.~\cite{Collins} as a measure of Lorentz symmetry 
breaking. Surprisingly, $\xi$ turns out to be fully independent of the 
intensity of the noncommutativity and, therefore, the calculated value 
of $\xi$ cannot be put under any experimental constraint. 
Hence, supersymmetry does not protect the noncommutative WZ model from the
Lorentz fine tuning problem.

For the basics of noncommutative quantum field theory we refer the 
reader to the paper by Minwalla et al.~\cite{Minwalla} as well as to the 
review articles~\cite{Douglas,Szabo,Gomes,Girotti1}. We work with the 
so-called canonical noncommmutavity, where time and position are 
considered to be self-adjoint operators ($q^{\mu}, \mu = 0, 1, 2, 3$) 
obeying the commutation algebra

\begin{equation}
\label{1}
\left[q^{\mu}\,,\,q^{\nu}\right]\,=\,i\,\Theta^{\mu \nu}\,.
\end{equation}

\noindent
Here, $\Theta^{\mu \nu}$ is an element of a real numerical 
antisymmetric matrix ($||\Theta ||$), parametrizing the noncommutativity. One can 
show that it is possible to retain the formulation of the theory in an 
ordinary (commutative) Minkowski space-time but deforming the ordinary 
product between field functions into the 
Gr\"onewold-Moyal~\cite{Gronewold,Moyal} ($*$) nonlocal product.

We shall be dealing here with noncommutative quantum field theories 
defined by means of a Lagrangian density in which all ordinary field 
products are replaced by Gr\"onewold-Moyal ones, and the corresponding 
Feynman rules are obtained through the usual functional integral methods. 
Thus, the Feynman propagators remain as in the commutative situation. 
Only the elementary vertices are modified by the 
noncommutativity~\cite{footnote4}. Observe that in the context of quantum gravity, as studied 
in~\cite{Collins,Vucetich,Jain}, the situation is exactly reversed: the 
propagators become modified while the vertices remain unchanged.

Already at the tree level, the presence of a constant matrix 
($||\Theta||$) has several consequences. Clearly, it breaks Lorentz invariance and, as a consequence,  the Lorentz algebra turns out to be deformed by terms proportional to $\Theta^{\mu \nu}$~\cite{Pengpan}. Some of the discrete symmetries are also broken, although PCT symmetry is preserved irrespective of the form of $||\Theta||$~\cite{Chaichian}. The prominent feature is that, at this level, the limit $\Theta^{\mu \nu} \rightarrow 0$ leads us back smoothly to the commutative situation. 

Our main concern in this paper is to determine the modifications that 
the radiative corrections introduce in the picture described above. This 
can only be done in connection with some specific model. Also, from now 
on, we assume $\Theta^{0j} = 0$ to evade unitarity and causality 
problems~\cite{Gomis,Seiberg,Girotti2}.

As we mentioned earlier, in this work we choose the noncommutative WZ 
model as a prototype of a fully consistent supersymmetric 
noncommutative theory~\cite{Girotti3,Bichl}. It is defined by the Lagrangian 
density

\begin{equation}
\label{7}
{\cal L} ={\cal L}_0\,+\,{\cal L}_m\,+\,{\cal L}_g\,,
\end{equation}

\noindent
where

\begin{subequations}
\label{8}
\begin{eqnarray}
&& {\cal L}_0\,=\,\frac12  A(-\partial^2)A+\frac12  B(-\partial^2)B
+ i \frac12 \overline \psi \not \! \partial \psi +\frac12 F^2+\frac12 
G^2\,,\label{a8}\\
&& {\cal L}_m\,=\, m F A + m G B - \frac{m}{2} \overline \psi 
\psi\,,\label{b8}\\ 
&& {\cal L}_g\,=\, g (F * A * A- F * B * B + G * A * B + G * B * A -
\overline \psi * \psi * A - 
\overline \psi *  i\gamma_5 \psi * B)\,.\label{c8}
\end{eqnarray}
\end{subequations}

\noindent
Here, $A$ is a scalar field, $B$ is a pseudo scalar field, $\psi$ is a 
Majorana spinor field, $F$ and $G$ are, respectively, scalar and 
pseudoscalar auxiliary fields and $g$ is the coupling 
constant~\cite{footnote2}.

We stress that, although Poincar\'e symmetry is partially broken by the 
noncommutativity, translations and supersymmetry generators still form 
a closed undeformed algebra,
i.e.,  supersymmetry is not affected by the presence of the tensor 
$\Theta^{\mu \nu}$~\cite{Chu:1999ij,Pengpan}. Indeed, one can verify that 
the action defined by~(\ref{7}) is still invariant under the 
supersymmetry transformations,

\begin{subequations}
\label{III-8}
\begin{eqnarray}
&&\delta A\,=\,\bar{\alpha}\,\psi\,,\label{aIII-7}\\
&&\delta B\,=\,-\,i\,\bar{\alpha}\,\gamma_5\,\psi\,,\label{bIII-7}\\
&&\delta \psi\,=\,-\,i\,\not \!\partial (A - i \gamma_5 
B)\,\alpha\,+\,(F - i \gamma_5\,G)\,\alpha\,,\label{cIII-7}\\
&&\delta F\,=\,-\,i\,\bar{\alpha}\,\not \!\partial 
\psi\,,\label{dIII-7}\\
&&\delta G\,=\,-\,\bar{\alpha}\,\gamma_5\,\not \!\partial 
\psi\,,\label{eIII-7}
\end{eqnarray}
\end{subequations}

\noindent
where $\alpha$ is a constant anticommuting Majorana spinor. The proof 
of invariance of the noncommutative WZ action under~(\ref{III-8}) is 
based on the observation that, for any fields $f$ and $g$,

\begin{equation}
\label{III-9}
\delta \left(f \ast g\right)\,=\,\delta f \ast g\,+\,f \ast \delta 
g\,.
\end{equation}

We focus now on the one loop corrections to the self-energy of the 
field $A$, to be hereafter denoted by $\Gamma\left( p \right)$.  This 
object was already computed in Ref.~\cite{Girotti3} and found to read

\begin{equation}
\label{9}
\Gamma\left( p \right)\,=\,-\,8\,i\, g^2\int \frac{d^4k}{(2\pi)^4 
}\,(p\cdot k)\,
\cos^2(k\wedge p) \Delta_F (k) \Delta_F (k+p)\,,
\end{equation}

\noindent
where $k \wedge p\,\equiv\,\frac{1}{2}\,k_{\mu} p_{\nu}\,\Theta^{\mu 
\nu}$ and 
$\Delta_F (k)$ is the free scalar propagator.

As usual, we split $\Gamma\left( p \right)$ into a planar,

\begin{equation}
\label{12}
\Gamma^{(P)}\left( p \right)\,=\,4\,i\,g^2\,\int \frac{d^4k}{(2 
\pi)^4}\,\frac{p \cdot k}{(k^2\,-\,m^2\,+\,i\e)\,\left[(k + p)^2\,-m^2\,+i\e 
\right]}\,,
\end{equation}

\noindent
and a nonplanar contribution,

\begin{equation}
\label{13}
\Gamma^{(NP)}\left( p \right)\,=\,4\,i\,g^2\,\int \frac{d^4k}{(2 
\pi)^4}\,\cos (2 p \wedge k)\,\frac{p \cdot 
k}{(k^2\,-\,m^2\,+\,i\e)\,\left[(k + p)^2\,-m^2\,+i\e \right]}\,.
\end{equation}

Power counting tell us that $\Gamma^{(P)}\left( p \right)$ is UV 
linearly divergent. However, the would be linearly divergent terms are washed 
out by symmetric integration. As for the remaining logarithmic UV 
divergences, one gets rid of them through a renormalization procedure that 
preserves Lorentz invariance in all stages of the calculation.

We shall focus on the nonplanar contribution, which contains the 
Lorentz non-invariant factor $\cos (2 p \wedge k)$. The integral in 
$\Gamma^{(NP)}\left( p \right)$ is UV finite and can be explicitly 
calculated~\cite{Gradstein}. One obtains

\begin{eqnarray}
\label{16}
&&\Gamma^{(NP)}\left( p \right)\,=\,\frac{g^2}{2 
\pi^2}\,p^2\,\int_0^1dx\,x\,K_0\left( \sqrt {a^2\,p \circ p}\right)\,,
\end{eqnarray}

\noindent
where $K_0(z)$ is the modified Bessel function of order zero, 
$a^2\,\equiv\,m^2\,-\,x(1 - x)\,p^2$, and

\begin{equation}
\label{20}
p \circ p\,=\,\left({\vec p} \cdot {\vec \theta} \right)^2\,-\,{\vec 
p}^{\,2}\,{\vec \theta}^{\,2}\,.
\end{equation}

\noindent
Here, we have introduced the three-vector $\theta^j,\,j = 1, 2, 3$, 
dual to the antisymmetric tensor $\Theta^{jk}$, namely,

\begin{equation}
\label{19}
\Theta_{jk}\,=\,\e_{jkl}\,\theta^l\,,
\end{equation}

\noindent
where $\e_{jkl}$ denote the covariant components of the fully 
antisymmetric Levi-Civit\`a tensor. 

According to Eq.~(\ref{20}), the quantity $p \circ p$ may tend to zero 
because either $\theta = |{\vec \theta}|\rightarrow 0$, or ${\vec p} 
\rightarrow 0$, or both. If we let $\theta \rightarrow 0$, keeping all 
components of the four-vector $p^{\mu}$ different from zero, 
$\Gamma^{(NP)}\left( p \right)$ develops a logarithmic singularity. However, if we 
carry out the same operation in the integrand of Eq.~(\ref{13}) we just 
recover the commutative counterpart of the two-point vertex function. 
Hence, the integral in Eq.~(\ref{13}) is not an analytic function of its 
integrand. This effect is well known to occur in noncommutative field 
theories and it is at the root of the UV/IR mechanism~\cite{Minwalla}. 
Of course, for $p^{\mu} \rightarrow 0$ and $\theta$ arbitrary one gets 
$\Gamma^{(NP)}\left( p \right) = 0$. In the case of quantum gravity the 
role of $\theta$ appears to be played by Planck's cutoff $1/\Lambda^2$. 
This analogy is based on the fact that these are the dimensional 
parameters that characterize the scale of Lorentz breaking in each case.

One could also consider the noncommutative theory with an UV cutoff $\Lambda_{UV}$, in which case there would be an additional scale to be taken into consideration. This has been done, for example, in Ref.~\cite{anisimov}. In this case, the amount of Lorentz violation found at the low energy level would depend on the relative strength of the scales of noncommutativity and $\Lambda_{UV}$, as discussed in~\cite{anisimov}. For the sake of simplicity, we will not introduce such cutoff, as our conclusions are insensitive to its presence.

We turn now into the problem of quantifying the breaking of Lorentz 
invariance. It was suggested in Ref. \cite{Collins} that the appropriate 
quantity for this purpose is 
$\xi \equiv \left[ \partial^2\Gamma^{(NP)}\left( p \right)/\partial (p^0)^2 + 
\partial^2\Gamma^{(NP)}\left( p \right)/\partial (p^1)^2 \right]_{p=0}$. 
However, for 
simplifying purposes, it is convenient to adopt the following generalization 
of this definition,

\begin{equation}
\label{22}
\xi\,=\,\left[ \frac{\partial^2\Gamma^{(NP)}\left( p \right)}{\partial 
(p^0)^2}\, + \,\frac{1}{3}\,\sum_{j = 1}^3 \frac{\partial^2\Gamma^{(NP)}\left( 
p \right)}{\partial (p^j)^2} \right]_{p=0} \,.
\end{equation} 

\noindent
By combining Eqs.~(\ref{16}) and (\ref{22}) one finds

\begin{equation}
\label{23}
\xi\,=\,\frac{g^2}{3 \pi^2}\,.
\end{equation}

\noindent
Unexpectedly, the breaking of Lorentz invariance does not depend on the 
intensity ($\theta$) of the noncommutativity. In particular, $\xi$ 
remains different from zero even at the limit $\theta \rightarrow 0$. 

It is instructive to compare this result with the corresponding one 
encountered by Collins et al. This essentially implies in replacing the 
ingredient of noncommutativity by an assumed quantum gravity effect. The 
model chosen as testing example in Ref.~\cite{Collins} was the Yukawa 
theory. The interaction Lagrangian is ${\cal L}_g^{Y} = g\,\phi\,{\bar 
\Psi}\,\Psi$, where $\phi$ is a real scalar field, while $\Psi$ is a 
Dirac field. In the standard case, the one loop contribution to the scalar 
field self energy ($\Pi(p)$) would be given by an UV quadratically 
divergent integral. However, it is assumed that this divergence is 
regulated by quantum gravity effects at very short distances, and these can be 
modeled by a cutoff function $f(x)$.
This function should verify $f(0) = 1$, while vanishing at infinity 
fast enough to make all Feynman integrals absolutely convergent. With 
these assumptions in mind, one is allowed to write

\begin{equation}
\label{25}
\Pi(p)\,=\,-\,4\,i\,g^2\,\int \frac{d^4k}{(2 
\pi)^4}\,f\left(\frac{|\vk|}{\Lambda}\right)\, f\left(\frac{|{\vec p} - 
\vk|}{\Lambda}\right)\,\frac{k^2\,-\,p \cdot k\,+\,m^2}{\left(k^2 - m^2 + 
i\e\right)\left[\left(k - p\right)^2\,-\,m^2\,+i\e\right]}\,,
\end{equation}

\noindent
where $\Lambda$ is a mass of the order of the Planck mass. From this 
expression, the authors of Ref.~\cite{Collins} obtain

\begin{equation}
\label{26}
\xi\,=\,\frac{g^2}{6 \pi^2}\left[1\,+\,2\,\int_0^{\infty} 
dx\,x\,f'(x)^2\right]\,.
\end{equation}

Leaving aside of consideration irrelevant numerical factors, the first 
term in the right hand side of Eq.~(\ref{26}) plays a role analog to 
that in the right hand side of Eq.~(\ref{23}). Indeed, neither $\theta 
\rightarrow 0$ nor $\Lambda^2 \rightarrow \infty$ enables one to recover 
Lorentz invariance. We see that in both cases the quantum corrections 
induce large Lorentz breaking contributions at low energies. The source 
of the problem in the model studied by us is, however, quite distinct 
from the one considered by Collins et al., since the degree of UV 
divergence of the starting loop integrals (see Eqs.~(\ref{13}) and 
(\ref{25})) is different.

We recall that, in the commutative situation, 
supersymmetry act as a custodial symmetry of Lorentz invariance at low 
energies, thus removing the need for a Lorentz fine tuning mechanism~\cite{Jain}. 
This happens because supersymmetry secures the cancellation of the leading - 
quadratic by power counting - contributions to the vertex functions of 
the quantum theory. The specific model used in Ref.~\cite{Jain} to 
exemplify this mechanism was, precisely, the  WZ model with a Lorentz 
violating cut-off function. On the other hand, we have presently used the WZ 
model to show that, when noncommutativity comes into play, 
supersymmetry is no longer powerful enough to protect Lorentz invariance at any 
energy scale. In the noncommutative situation, it is the non-analyticity 
induced by the Gr\"onewold-Moyal product which is responsible for the 
large violation of Lorentz invariance at low energies.

We close this note by recalling some phenomenological 
studies~\cite{anisimov,Carlson:2001sw,Carlson:2002zb}, where it has been pointed out the 
existence of difficulties to conciliate canonical noncommutativity 
with available data on Lorentz symmetry breaking in low-energy 
experiments. This is due to the generation of Lorentz-violating operators in the 
noncommutative versions of QED and QCD, like, for instance, the 
coupling of fermions with a background magnetic field depending on $||\Theta
||$. However, such theories have problems of their own. In fact, both of 
them are plagued by quadratic infrared divergences~\cite{Hay} arising 
from the UV/IR mechanism which, as asserted in Ref.~\cite{Minwalla}, 
invalidate their perturbative expansions. In this connection, the novelty 
would be to find some inconsistency with Lorentz invariance in a well 
defined and realistic noncommutative gauge theory. Up to our knowledge, 
only supersymmetric noncommutative gauge theories have proved to be 
fully consistent, at least in the one-loop approximation~\cite{Ferrari1}. 
To provide further back up for this line of argument, we recall that 
the dangerous Lorentz violating operators mentioned in the last paragraph 
of Ref.~\cite{Carlson:2002zb} do not show up in the supersymmetric 
version of noncommutative QED, if supersymmetry is left unbroken. Of 
course, it makes no sense to try to conciliate the outcomes from an 
extremely simple theory, like the noncommutative WZ model, with experimental 
data.

We can also entertain the hope that alternative approaches to 
space-time  noncommutativity, like the one suggesting the introduction of the 
tensor $\Theta^{\mu \nu}$ in a Lorentz covariant 
way~\cite{Carlson:2002wj,Morita:2002cv}, may improve the agreement with currently available 
data. In any case, whether or not we can conciliate space-time 
noncommutativity with the absence of Lorentz violation in available low energy 
experiments is a very interesting question which certainly deserves 
further study.

\newpage


{\bf Acknowledgements.} A. F. Ferrari is thankful to Prof. V. O. 
Rivelles for discussions. H. O. Girotti and A. F. Ferrari are indebted to F. 
S. Bemfica for very useful remarks. This work was partially supported 
by Conselho Nacional de Desenvolvimento Cient\'\i fico e Tecnol\'ogico 
(CNPq) and Funda\c{c}\~ao de Amparo \`a Pesquisa do Estado de S\~ao 
Paulo (FAPESP). H. O. Girotti acknowledges support from PRONEX under 
contract CNPq 66.2002/1998-99. A. F. Ferrari has been supported by FAPESP, 
project No. 04/13314-4.



\begin{thebibliography}{99}

\bibitem {Collins} J. Collins, A. Perez, D. Sudarsky, L. Urrutia, H. 
Vucetich, Phys. Rev. Lett. {\bf 93}, 191301 (2004).

\bibitem{anisimov} A.~Anisimov, T.~Banks, M.~Dine and M.~Graesser, 
Phys.\ Rev.\ D {\bf 65}, 085032 (2002).

\bibitem {Jain} P. Jain, J. P. Ralston, Phys. Lett. {\bf B621}, 213 
(2005).

\bibitem{Mat} A. Matusis, L. Susskind, N. Toumbas, JHEP {\bf 12}, 002 (2000).

\bibitem {Girotti3} H. O. Girotti, M. Gomes, V. O. Rivelles, A. J. da 
Silva, Nucl. Phys. {\bf B587}, 299 (2000).

\bibitem{Bichl} A.A. Bichl, M. Ertl, A. Gerhold, J.M. Grimstrup, H. 
Grosse, L. Popp, V. Putz, M. Schweda, R. Wulkenhaar, Int.\ J.\ Mod.\ 
Phys.\ A {\bf 19}, 4231 (2004).

\bibitem{Minwalla} S. Minwalla, M. van Raamsdonk, N. Seiberg, JHEP 
{\bf
02}, 020 (2000).

\bibitem{Douglas} M. Douglas, N. A. Nekrasov, Rev. Mod. Phys. {\bf 
73}, 977 (2001).

\bibitem{Szabo} R. Szabo, Phys. Rept. {\bf 378}, 207 (2003). 

\bibitem{Gomes} M. Gomes in  Proceedings of the XI Jorge Andr\'e
Swieca Summer School, Particles and Fields, G. A. Alves,
O. J. P. \'Eboli and V. O. Rivelles eds, World Scientific Pub. Co,
2002. 

\bibitem{Girotti1} H. O. Girotti, ``Noncommutative Quantum Field 
Theories'',
hep-th/0301237.

\bibitem{Gronewold} H. J. Gr\"onewold, Physica {\bf 12}, 405-460 (1946).

\bibitem {Moyal} J. E. Moyal, Proc. Cambridge Philos. Soc. {\bf 45}, 99-124 (1949).

\bibitem {footnote4} See, for instance, 
Refs.\cite{Minwalla,Douglas,Szabo,Gomes,Girotti1}. 

\bibitem {Vucetich} H. Vucetich, ``Testing Lorentz Invariance Violation 
in Quantum Gravity Theories'', gr-qc/0502093.

\bibitem {Pengpan} T. Pengpan, X. Xiong, Phys. Rev. {\bf D63}, 085012 
(2001). 

\bibitem {Chaichian} M. Chaichian, K. Nishijima, A. Turenu, Phys. Latt. 
{\bf B568}, 146 (2003).

\bibitem {Gomis} J. Gomis, T. Mehen, Nucl. Phys. {\bf B591}, 265 
(2000).

\bibitem {Seiberg} N. Seiberg, L. Susskind, N. Toumbas, JHEP {\bf0006}, 
044 (2000).

\bibitem {Girotti2} H. O. Girotti, M. Gomes, A. Yu. Petrov, V. O. 
Rivelles, A. J. da Silva, JHEP {\bf0205}, 040 (2002).

\bibitem {footnote2} Our Minkowskian metric is $g^{00} = - g^{11} = - 
g^{22} = - g^{33} = + 1$. Furthermore, we use Dirac's representation for 
the $\gamma$ matrices and $\gamma_5 \equiv i \gamma^0 \gamma^1 \gamma^2 
\gamma^3$ implying that $\gamma_5^{\dagger} = \gamma_5$ and $\gamma_5^2 
= 1$.

\bibitem{Chu:1999ij} { Among several references dealing with this 
subject we single out} C.~S.~Chu and F.~Zamora, ``Manifest supersymmetry in non-commutative geometry'', JHEP {\bf 0002}, 022 (2000).

\bibitem {Gradstein} I. M. Gradshteyn and I. M. Ryzhik, {\it Tables of 
integrals, Series, and Products}, (Academic Press, Inc., New York, 
1980).

\bibitem{Carlson:2001sw} C.~E.~Carlson, C.~D.~Carone and R.~F.~Lebed, ``Bounding noncommutative QCD'', Phys.\ Lett.\ B {\bf 518}, 201 (2001).

\bibitem{Carlson:2002zb}
  C.~E.~Carlson, C.~D.~Carone and R.~F.~Lebed,
  Phys.\ Lett.\ B {\bf 549}, 337 (2002).

\bibitem{Hay} M. Hayakawa. Phys. Lett. {\bf B478}, 394 (2000);
``Perturbative analysis of infrared and ultraviolet aspects of
noncommutative QED on $R^4$'', hep-th/9912167.

\bibitem {Ferrari1} 
A. F. Ferrari, H. O. Girotti, M. Gomes, A. Yu. Petrov, A. A. Ribeiro, V. O. Rivelles, A. J. da Silva, 
Phys. Rev. {\bf D69}, 025008 (2004); 
Phys. Rev. {\bf D70}, 085012 (2004),  
A. F. Ferrari, H. O. Girotti, M. Gomes, A. Yu. Petrov, A. A. Ribeiro, A. J. da Silva, 
Phys. Lett {B577}, 83 (2003); 
Phys. Lett {B601}, 88 (2004).

\bibitem{Carlson:2002wj}
  C.~E.~Carlson, C.~D.~Carone and N.~Zobin,
  Phys.\ Rev.\ D {\bf 66}, 075001 (2002).

\bibitem{Morita:2002cv}
  K.~Morita,
  Prog.\ Theor.\ Phys.\  {\bf 108}, 1099 (2002).


\end{thebibliography}
\end{document}